\documentclass[twocolumn,aps,prd,10pt,superscriptaddress,longbibliography]{revtex4-1}

\usepackage{amssymb,amsmath}
\usepackage{bm}
\usepackage{dcolumn}
\usepackage{glossaries}
\usepackage{graphicx}
\usepackage[version=4]{mhchem}
\usepackage[usenames,dvipsnames,svgnames]{xcolor}
\usepackage{siunitx}
\usepackage{hyperref}
\usepackage{makecell}
\hypersetup{
pdfnewwindow=true,      
colorlinks=true,        
linkcolor=Blue,         
citecolor=Blue,         
filecolor=Blue,         
urlcolor=Blue           
}

\newacronym{ace}{ACE}{atomic cluster expansion}
\newacronym{md}{MD}{molecular dynamics}
\newacronym{gpu}{GPU}{graphic processing units}
\newacronym{mlp}{MLP}{machine-learned potential}
\newacronym{nep}{NEP}{neuroevolution potential}
\newacronym{2d}{2D}{two-dimensional}
\newacronym{dft}{DFT}{density functional theory}
\newacronym{apcvd}{AP-CVD}{atmosphericpressure chemical vapor deposition}
\newacronym{hnemd}{HNEMD}{homogeneous non-equilibrium molecular dynamics}
\newacronym{mfp}{MFP}{mean free path}
\newacronym{emd}{EMD}{equilibrium molecular dynamics}
\newacronym{shc}{SHC}{spectral heat current}
\newacronym{nemd}{NEMD}{non-equilibrium molecular dynamics}
\newacronym{dfpt}{DFPT}{density functional perturbation theory}
\newacronym{sw}{SW}{Stillinger-Weber}
\newacronym{vasp}{VASP}{Vienna \emph{ab initio} simulation package}
\newacronym{rmse}{RMSE}{root mean square error}
\newacronym{aimd}{AIMD}{\emph{ab initio} molecular dynamics}
\newacronym{ann}{ANN}{artificial neural network}
\newacronym{ltc}{LTC}{lattice thermal conductivity}
\newacronym{nn}{NN}{neural network}
\newacronym{snes}{SNES}{separable natural evolution strategy}
\newacronym{bte}{BTE}{Boltzmann transport equation}
\newacronym{ald}{ALD}{anharmonic lattice dynamics}

\DeclareSIUnit\angstrom{\text{Å}}
\DeclareSIUnit{\atom}{atom}
\DeclareSIUnit{\step}{step}
\DeclareSIUnit{\atomstepsecond}{\atom\step\per\second}

\begin{document}

\title{Phonon coherence and minimum thermal conductivity in disordered superlattice}

\author{Xin Wu}
\affiliation{Institute of Industrial Science, The University of Tokyo, Tokyo 153-8505, Japan}

\author{Zhang Wu}
\affiliation{AVIC Jiangxi Hongdu Aviation Industry Group Company Ltd., Nanchang 330024, P. R. China}

\author{Ting Liang}
\affiliation{Department of Electronic Engineering and Materials Science and Technology Research Center, The Chinese University of Hong Kong, Shatin, N.T., Hong Kong SAR, 999077, P. R. China}

\author{Zheyong Fan}
\email{brucenju@gmail.com}
\affiliation{College of Physical Science and Technology, Bohai University, Jinzhou 121013, P. R. China}

\author{Jianbin Xu}
\affiliation{Department of Electronic Engineering and Materials Science and Technology Research Center, The Chinese University of Hong Kong, Shatin, N.T., Hong Kong SAR, 999077, P. R. China}

\author{Masahiro Nomura}
\email{nomura@iis.u-tokyo.ac.jp}
\affiliation{Institute of Industrial Science, The University of Tokyo, Tokyo 153-8505, Japan}

\author{Penghua Ying}
\email{hityingph@tauex.tau,ac.il}
\affiliation{Department of Physical Chemistry, School of Chemistry, Tel Aviv University, Tel Aviv, 6997801, Israel}

\date{\today}

\begin{abstract}

Phonon coherence elucidates the propagation and interaction of phonon quantum states within superlattice, unveiling the wave-like nature and collective behaviors of phonons. 
Taking MoSe$_2$/WSe$_2$ lateral heterostructures as a model system, we demonstrate that the intricate interplay between wave-like and particle-like phonons, previously observed in perfect superlattice only, also occurs in disordered superlattice. 
By employing molecular dynamics simulation based on a highly accurate and efficient machine-learned potential constructed herein, we observe a non-monotonic dependence of the lattice thermal conductivity on the interface density in both perfect and disordered superlattice, with a global minimum occurring at relatively higher interface density for disordered superlattice. 
The counter-intuitive phonon coherence contribution can be characterized by the lagged self-similarity of the structural sequences in the disordered superlattice. 
Our findings extend the realm of coherent phonon transport from perfect superlattice to more general structures, which offers more flexibility in tuning thermal transport in superlattices.

\end{abstract}
\maketitle
\raggedbottom

\section{Introduction}

Phonon thermal transport exhibits significant potential in nanoscale thermal physics, especially in semiconductors and insulators where the \gls{ltc} is almost entirely derived from lattice vibrations \cite{chen2005nanoscale}. As the quanta of lattice vibrations, phonons exhibit particle-like behavior through energy quantization and quasi-particle interactions, playing a crucial role in thermal transport by affecting \gls{ltc} through scattering and collision processes. However, the wave nature of phonons is equally significant \cite{Zhang_2022_How, Zhang_2021_Generalized}. It describes the characteristics of lattice vibrations through the relationship between wave vector and frequency, manifesting as dispersion relations in wave equations. By controlling phonon wave vectors and frequencies, one can influence phonon propagation paths and scattering mechanisms. For instance, when the phonon wavelength is comparable to the characteristic dimensions of a material, phonon transport exhibits significant coherence \cite{Tamura_1988_Acousticphonon, Simkin_2000_Minimuma, Luckyanova_2012_Coherenta, Latour_2014_Microscopic, Felix_2018_Thermal, Wu_2022_Transition, Nomura_2022_Review}. This means phonons can travel coherently without losing phase information, thereby significantly altering the \gls{ltc} and thermal transport properties of the material.

In periodically alternating nanostructures known as superlattices, composed of two or more different materials, phonons experience multiple reflections and interference at periodic interfaces, leading to the formation of new phonon spectra and coherent transport \cite{Colvard_1985_Folded, Tamura_1988_Acousticphonon}. The wave nature of phonons becomes critically important. Perfect superlattices provide an ideal platform to explore and utilize the coherent transport properties of phonons \cite{Hu_2020_MachineLearningOptimizeda, Wang_2014_Decomposition, Wei_2022_Quantifying}. This behavior has been theoretically predicted and experimentally verified in various systems, from semiconductors such as GaAs/AlAs \cite{Luckyanova_2012_Coherenta, Hu_2020_MachineLearningOptimizeda} to \gls{2d} materials like graphene/$h$-BN \cite{Felix_2018_Thermal, Wu_2024_Unexpected, Wu_2022_Transition}. However, disordered superlattices offer complex interface scattering and localization effects, making them ideal systems to contrast with perfect superlattices. These structures are often used to verify phonon coherence in superlattices but have not received equivalent attention or treatment in their own right and remain elusive.

For predicting the properties of new structures and materials, \gls{md} simulation is an excellent choice, which can implicitly include all orders of lattice anharmonicity and phonon scattering and provide detailed atomic-level information \cite{Bao_2018_Review}. It circumvents the limitations of \gls{bte} in describing higher-order phonon scattering and in handling complex systems with large unitcells. Additionally, \gls{mlp} \cite{Behler_2016_Perspective} offers high-precision descriptions of interatomic interactions, providing excellent descriptions for thermal transport in many materials \cite{Wang_2023_Quantumcorrected, Liang_2023_Mechanismsa, ying2023sub, Xu_2023_Accuratea, Dong_2024_Molecular}.

\begin{figure*}[htb]
\begin{center}
\includegraphics[width=1.5\columnwidth]{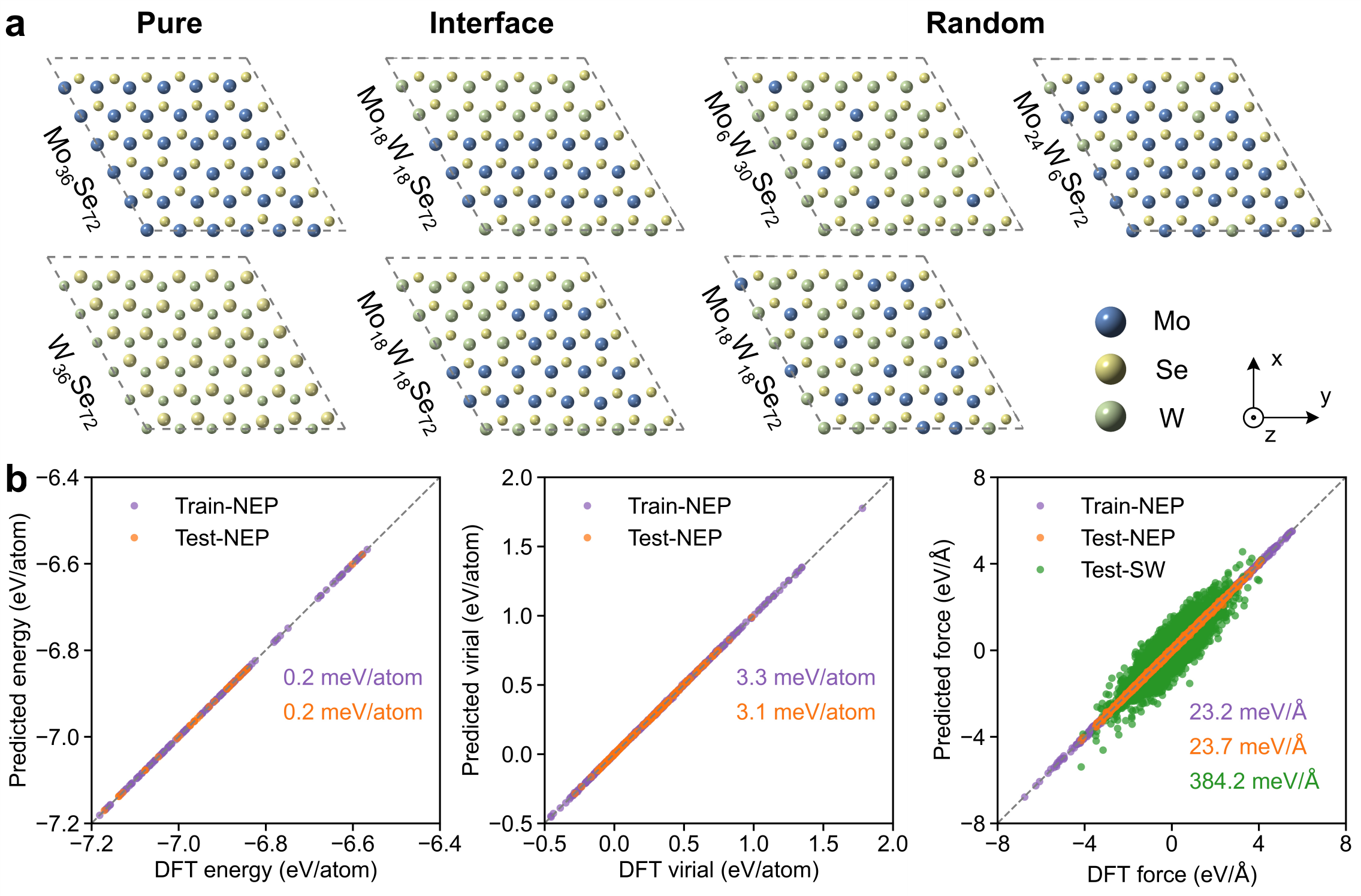}
\caption{The construction of NEP model. (a) The cells of the reference structures, including pure MoSe$_2$, pure WSe$_2$, and their heterostructures. (b) The energy, virial, and force calculated from the NEP model and the SW potential compared to the DFT reference data.}
\label{fig:nep}
\end{center}
\end{figure*}

In this work, we developed a \gls{mlp} based on the \gls{nep} framework \cite{Fan_2021_Neuroevolution, Fan_2022_Improvinga,Fan_2022_GPUMD} and used nearly lattice-matched \gls{2d} MoSe$_2$/WSe$_2$ (2H-phase) lateral heterostructures as the model systems. We investigated the \gls{ltc} of perfect and disordered superlattices at room temperature with various interface densities. Detailed analyses of the spectral \gls{ltc} elucidate the origins of two distinct minimum thermal conductivities in perfect and disordered superlattices. We also established a factor of disorder to describe the \gls{ltc} trends in disordered superlattices and revealed the \gls{ltc} variations in the coherent transport regime of perfect superlattices through phonon band folding. 

\section{Results and discussions}
\subsection{NEP model evaluation}
\label{section:mlp}
\begin{figure*}[htb]
\begin{center}
\includegraphics[width=1.5\columnwidth]{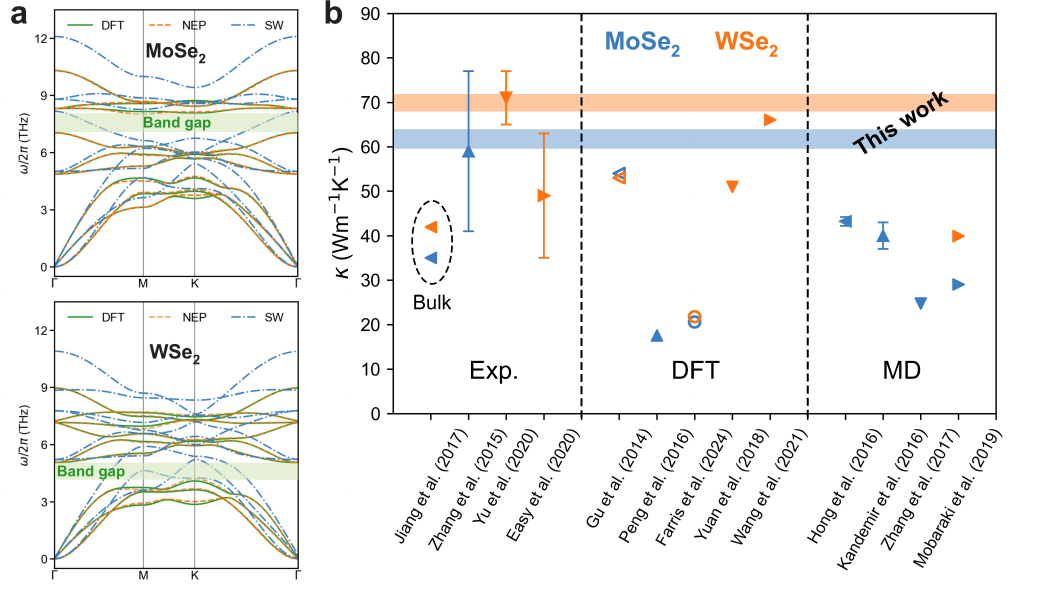}
\caption{Benchmark of the accuracy of the \gls{nep} model for MoSe$_2$ and WSe$_2$. (a) The phonon dispersion relations of MoSe$_2$ and WSe$_2$ calculated by \gls{dft}, the \gls{nep} model, and the SW potential. (b) Comparison of \gls{ltc} of monolayer MoSe$_2$ and WSe$_2$ with respect to other experimental \cite{Jiang_2017_Probing, Zhang_2015_Measurement, Yu_2020_InPlane, Easy_2021_Experimental}, DFT-based \cite{Gu_2014_Phonon, Peng_2016_Thermal, Farris_2024_Microscopic, Yuan_2018_Effects, Wang_2021_Role}, and SW-based MD \cite{Hong_2016_Thermal, Kandemir_2016_Thermal, Zhang_2017_Phonon, Mobaraki_2019_Temperaturedependent} results at the room temperature. The two horizontal bands are the results with error bars from this work by \gls{nep}-based \gls{md}, which show good agreement with the experiments.}
\label{fig:benchmark}
\end{center}
\end{figure*}

We employed the \gls{nep} approach \cite{Fan_2021_Neuroevolution,Fan_2022_Improvinga,Fan_2022_GPUMD} (see \autoref{section:nep} for details) to train a \gls{mlp} for MoSe$_2$/WSe$_2$ lateral heterostructures against the total energy, atomic forces, and virials of a training dataset from \gls{dft} calculations (see \autoref{section:dft} for details). As shown in \autoref{fig:benchmark}(a), the training dataset includes pure MX$_2$ structures, their lateral heterostructures with both flat and embedded interfaces, and random ternary Mo$_x$W$_{1-x}$Se structures (see \autoref{section:dataset} for details).

As depicted in \autoref{fig:nep}(b), the \gls{nep} model achieves high accuracy in both training and test datasets, with \glspl{rmse} of energy, virial, and force below \qty{0.5}{\milli\electronvolt\per\atom}, \qty{3.3}{\milli\electronvolt\per\atom}, and \qty{23.7}{\milli\electronvolt\per\angstrom}, respectively. For comparison, we evaluated the atomic forces predicted by a \gls{sw} potential \cite{jiang2019misfit} against \gls{dft} results and found that its \gls{rmse} (\qty{384.2}{\milli\electronvolt\per\angstrom}) is an order of magnitude higher. In terms of computational efficiency, on a single GeForce RTX 4090 \gls{gpu}, the present \gls{nep} model can achieve 2.4 $\times$ 10$^{7}$ atom-steps per second as implemented in \textsc{GPUMD}.\cite{Fan_2022_GPUMD} This performance is comparable to the speed of the traditional \gls{sw} potential implemented in \textsc{LAMMPS} \cite{thompson2022lammps}, which achieves 5.8 $\times$ 10$^{7}$ atom-step per second using 64 Xeon Platinum 9242 CPU cores.

After obtaining the \gls{nep} model, we verified its accuracy by comparing the predicted phonon dispersion and \gls{ltc} of MoSe$_2$ and WSe$_2$ with results from \gls{dft}, \gls{sw}, as well as experimental measurement. As shown in \autoref{fig:benchmark}(a), the phonon dispersion curves predicted by the \gls{nep} model align perfectly with \gls{dft}, providing a much more accurate band gap for both MoSe$_2$ and WSe$_2$ compared to the \gls{sw} potential \cite{Jiang_2017_Handbook}. For the \gls{ltc} predictions for monolayer MoSe$_2$ and WSe$_2$ (see \autoref{fig:benchmark}(d)), our \gls{nep}-based \gls{hnemd} results (see \autoref{section:hnemd} for details) closely match experimental measurements \cite{Zhang_2015_Measurement, Yu_2020_InPlane, Easy_2021_Experimental}, whereas \gls{sw}-driven \gls{md} simulations \cite{Hong_2016_Thermal, Kandemir_2016_Thermal, Zhang_2017_Phonon, Mobaraki_2019_Temperaturedependent} significantly underestimate the \glspl{ltc}, aligning more with measured \gls{ltc} of bulk MoSe$_2$ and WSe$_2$ \cite{Jiang_2017_Probing}. The \gls{bte}-\gls{ald} predictions \cite{Gu_2014_Phonon, Peng_2016_Thermal, Farris_2024_Microscopic, Yuan_2018_Effects, Wang_2021_Role} based on \gls{dft}  force constants (labeled as ``DFT" in \autoref{fig:benchmark}(b)) exhibit large variations and less satisfactory alignment to experimental results than the predictions from \gls{nep}.

\subsection{The model lateral heterostructures}
\label{section:structure}

\begin{figure*}[htb]
\begin{center}
\includegraphics[width=1.5\columnwidth]{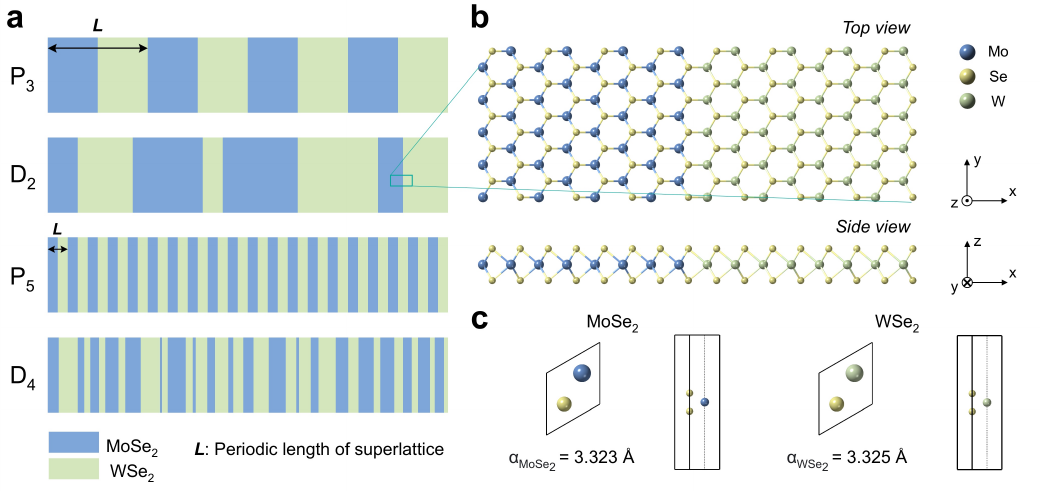}
\caption{Illustration of MoSe$_2$/WSe$_2$ lateral heterostructures. (a) Two pairs of MoSe$_2$/WSe$_2$ lateral heterostructures, each consisting of a perfect (P) and a disordered (D) superlattice with equal interface density: P$_3$ paired with D$_2$, and P$_5$ paired with D$_4$. $L$ denotes the periodic length of the superlattice. (b) Local atomic-level zoomed-in view of the MoSe$_2$/WSe$_2$ heterostructure interface. (c) The unit cell of MoSe$_2$ and WSe$_2$ and their lattice constants.}
\label{fig:model}
\end{center}
\end{figure*}

As shown in \autoref{fig:model}, we chose MoSe$_2$ and WSe$_2$ to form a binary model system, both in the 2H phase. The optimized lattice constant from  \gls{dft} (\gls{nep})  are 3.323 \AA\ and 3.325 \AA (3.321 \AA\ and 3.321 \AA) for MoSe$_2$ and WSe$_2$, respectively, resulting in a negligible lattice mismatch. Notably, a monolayer MoSe$_2$/WSe$_2$ lateral heterostructure was recently synthesized and found to exhibit electrical rectification \cite{Zhang_2022_Simultaneous}.

\begin{table}
\begin{center}
\caption{The MoSe$_2$/WSe$_2$ perfect and disordered superlattice with different interface densities considered in this work. The fourth column, $L$ denotes the periodic length of the superlattice (see \autoref{fig:model}(a))}
    \label{table:density}
    \setlength{\tabcolsep}{5pt}
    \begin{tabular}{cccc}
        \hline
        \makecell{Interface density\\ (nm$^{-1}$)} & Perfect   & Disordered  & \makecell{$L$ (nm)} \\ 
        \hline
        0.0351 & P$_1$ & N/A & 56.9500 \\ 
        0.0702 & P$_2$ & D$_1$ & 28.4750 \\ 
        0.1405 & P$_3$ & D$_2$ & 14.2370 \\ 
        0.3512 & P$_4$ & D$_3$ & 5.6950 \\ 
        0.7024 & P$_5$ & D$_4$ & 2.8470 \\ 
        0.8780 & P$_6$ & D$_5$ & 2.2780 \\ 
        1.2291 & N/A & D$_6$ & N/A \\ 
        1.7559 & P$_7$ & D$_7$ & 1.1390 \\ 
        2.1071 & N/A & D$_8$ & N/A \\ 
        2.6339 & N/A & D$_9$ & N/A \\ 
        3.1607 & N/A & D$_{10}$ & N/A \\ 
        3.5119 & P$_8$ & N/A & 0.5695 \\ 
        \hline
\end{tabular}
\end{center}
\end{table}

The lateral heterostructures here include two types: eight perfect superlattices (labeled as P$_1$$\sim$P$_8$) and ten disordered superlattices (labeled as D$_1$$\sim$D$_{10}$). For each superlattice, we define the interface density as the number of heterointerfaces per unit lateral length. This definition allows for a clear correspondence between perfect and disordered superlattices. The structural parameters of considered superlattices are listed in\autoref{table:density}, and \autoref{fig:model}(a) presents the sequences of two pairs of perfect and disordered superlattices ($P_3$ and $D_2$, $P_5$ and $D_4$) with identical interface density. For a disordered superlattice with the specified interface density, we generate the corresponding configurations using genetic algorithms \cite{whitley1994genetic} under a constraint of MoSe$_2$:WSe$_2$ = 1:1. In this case, the difference between perfect and disordered superlattices with the same interface density lies solely in the varying positions of the heterointerfaces.

One should note that for perfect superlattices, P$_1$ and P$_8$, which have the largest and smallest periodic lengths (see \autoref{table:density}) respectively, there are no corresponding disordered counterparts. In the case of P$_1$, with minimal interface density and only a single heterointerface, altering the position of the heterointerface would inevitably break the constraint of MoSe$_2$:WSe$_2$ = 1:1. For P$_8$ with maximum interface density, the periodic length equal to the lattice constant of component orthogonal unit cells. Thus, its all possible hetero-interface locations are fully occupied, leaving no room for further adjustment. Therefore, P$_1$ and P$_8$ represent the boundary limits for disordered superlattices.

\subsection{Thermal transport properties}
\label{section:kappa}

Among \gls{md} methods for predicting \gls{ltc} of \gls{2d} materials, the \gls{hnemd} method is widely used for its efficiency over other approaches such as \gls{emd} and \gls{nemd} \cite{Fan_2019_Homogeneousa, gu2021thermal, wu2021thermal, ying2023sub, dong2024molecular}. Therefore, we apply the \gls{hnemd} approach (see \autoref{section:hnemd} for details) to calculate the \gls{ltc} of MoSe$_2$/WSe$_2$ lateral heterostructures.

\begin{figure}[htb]
\begin{center}
\includegraphics[width=1\columnwidth]{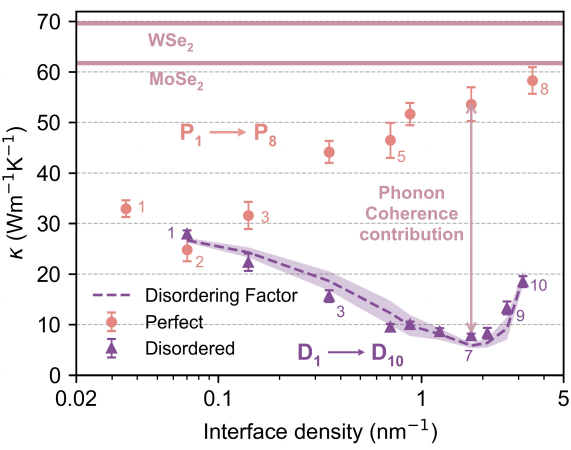}
\caption{Thermal conductivity of MoSe$_2$/WSe$_2$ lateral heterostructures, including perfect (P$_{1}$$\sim$P$_{8}$, filled circles) and disordered (D$_{1}$$\sim$D$_{10}$, filled triangles) superlattices, as a function of interface density at \qty{300}{\kelvin}. The two horizontal solid lines at the top represent the \gls{ltc} of monolayer MoSe$_2$ and WSe$_2$, respectively. The dashed line for disordered superlattices corresponds to the disordering factor $R$-based predictions using Eq.~\eqref{equation:disorder1} and Eq.~\eqref{equation:disorder2}. Groups marked with numbers will be selected as representative groups for subsequent phonon analysis in \autoref{fig:spectral}.}
\label{fig:TC}
\end{center}
\end{figure}

\autoref{fig:TC} presents the \gls{ltc} of MoSe$_2$/WSe$_2$ lateral heterostructures with different interface densities at the room temperature of \qty{300}{\kelvin}. As the interface density increases, the \gls{hnemd}-predicted \gls{ltc} of the perfect superlattice exhibits a non-monotonic trend, first decreasing and then increasing, with a global minimum at P$_2$ with an interface density of around \qty{0.07}{\nano\meter^{-1}}. This behavior reflects the transition of phonon transport from the incoherent to the coherent regime, a phenomenon previously reported in various superlattice systems \cite{Simkin_2000_Minimuma, Luckyanova_2012_Coherenta, Felix_2018_Thermal, Hu_2020_MachineLearningOptimizeda, Wu_2022_Transition}, which can be understood from the competition between \gls{ltc} reduction due to enhanced interface scattering and the \gls{ltc} increase driven by coherent phonon transport. In the incoherent transport regime of low interface density (below \qty{0.07}{\nano\meter^{-1}}), the \gls{ltc} of the perfect superlattice is primarily influenced by interface scattering, with the lower interface density of P$_1$ resulting in a significantly higher \gls{ltc} compared to P$_2$. Conversely, when the interface density is high (above \qty{0.07}{\nano\meter^{-1}}), phonon transport of perfect superlattices transitions the coherent regime, resulting in a monotonic increase in \gls{ltc} with interface density. Notably, at the maximum interface density of \qty{3.51}{\nano\meter^{-1}}, the superlattice P$_8$, with its smallest indivisible unit, reaches a maximum \gls{ltc} of \qty{58.32}{\watt\per\meter\per\kelvin}, approaching that of MoSe$_2$ (\qty{61.75}{\watt\per\meter\per\kelvin}) and WSe$_2$ (\qty{69.65}{\watt\per\meter\per\kelvin}), indicating that phonon coherence dominates \gls{ltc} in this region.

\begin{figure}[htb]
\begin{center}
\includegraphics[width=1\columnwidth]{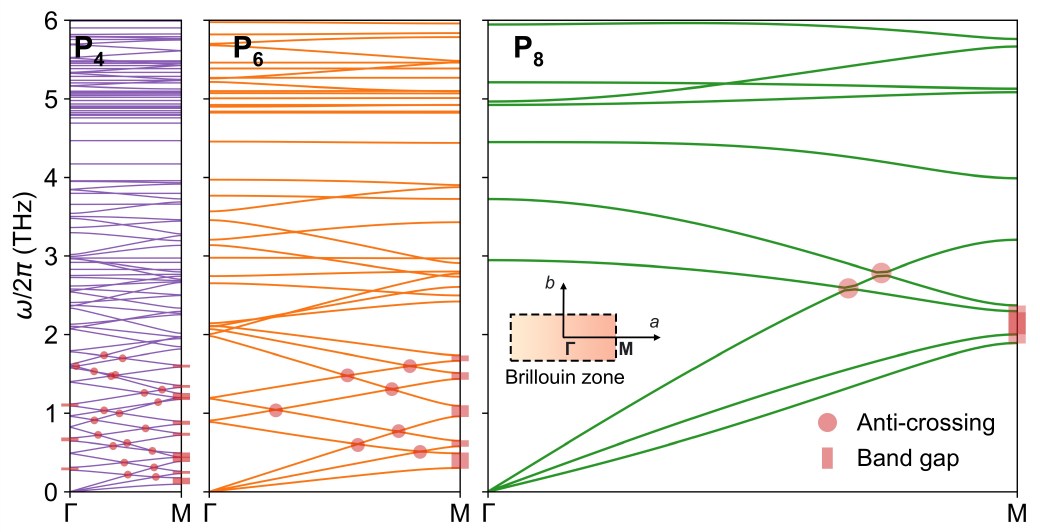}
\caption{Phonon dispersion relation of MoSe$_2$/WSe$_2$ lateral superlattice P$_4$, P$_6$, and P$_8$ within the coherent transport regime. The first Brillouin zone is defined by the smallest repeating unit of the superlattice, with the $\Gamma$-M direction is along the periodicity of its components, also the thermal transport direction.  The phonon dispersion results were calculated by the \textsc{calorine} \cite{Lindgren_2024_Calorine} together with \textsc{phonopy} packages \cite{togo2023first} based on the finite difference method, where the force constants were calculated from the \gls{nep} model.}
\label{fig:phonon}
\end{center}
\end{figure}

To further understand the phonon coherence observed in perfect superlattices, in \autoref{fig:phonon} we examined the phonon dispersion for P$_{4}$, P$_{6}$, and P$_{8}$ in the coherent transport regime. The clear band folding of acoustic phonons propagating through superlattice is observed here, allowing phonons to propagate in homogeneous materials and free of interfaces. However, this folding introduces two additional stop bands: One is the anti-crossing point, which refers to the intra-mode stop bands generated by the crossing of folded phonon branches with other phonons; another one is the band gap that occurred at the center and boundary of the folded Brillouin zone. As interface density increases (P$4$$\rightarrow$P${8}$), fewer bands in the perfect superlattice fold, reducing the number of stop bands, which may result in higher group velocity and phonon \gls{mfp}, thereby enhancing the \gls{ltc}.

In disordered superlattices, phonon coherence is expected to be nearly absent, as shown in previous studies \cite{Hu_2020_MachineLearningOptimizeda, Luckyanova_2012_Coherenta}. Thus, the difference in \gls{ltc} between the perfect and disordered superlattices with identical interface densities reflects the contribution of phonon coherence (see the vertical arrow in \autoref{fig:TC}). In this case, one may intuitively expect the \gls{ltc} of disordered superlattices to decrease monotonically with increasing interface density due to enhanced interface scattering. However, we observed a similar non-monotonic trend in \gls{ltc} for disordered superlattices, as seen in perfect superlattices, with the minimum \gls{ltc} occurring near the maximum interface density limit. This indicates that at very high interface densities (above \qty{1.76}{\nano\meter^{-1}}), wave-like phonons contribute significantly more to \gls{ltc} than particle-like phonons. We attribute this to the fact that, as the interface density approaches the maximum limit, the disordered superlattice begins to resemble a perfect superlattice, exhibiting a high degree of lagged self-similarity in the structural sequence. 

To quantify the lagged self-similarity for disordered superlattices, we can define a disordering factor that measures system disorder using the auto-correlation function of parameterized sequences as follows:
\begin{equation}
    R(\tau) =  \frac{1}{N-\tau} \sum^{N-\tau-1}_{t=0} \biggl(x(t)-\mu\biggl) \biggl( x(t+\tau)-\mu \biggl),
\label{equation:disorder1}
\end{equation}
where $\tau$ is the lag parameter, indicating the step number of sequence move backward, and $\mu$ is the average of parameterized sequences. The disordering factor $R$ can be further obtained by considering the average of different lag parameters $R(\tau$):
\begin{equation}
    R = \frac{1}{n}\sum^{n}_{\tau=1} R(\tau),
\label{equation:disorder2}
\end{equation}
Here we chose $n=4$ to quantify $R$. Furthermore, one can predict the \gls{ltc} using a single scaling factor $\alpha$ and a basis $\beta$: $\kappa = \alpha R + \beta$. Remarkably, the $\alpha R + b$ profiles are in excellent agreement with the \gls{ltc} of disordered superlattices predicted by \gls{hnemd} using $\alpha=$ \qty{20.72}{\watt\per\meter\per\kelvin} and $\beta =$ \qty{6.2}{\watt\per\meter\per\kelvin} (see \autoref{fig:TC}). This strong correlation suggests that the disordering factor serves as a simple and cost-effective predictor for interface density in disordered superlattices with minimum \gls{ltc}, and can be adapted to describe other aperiodic heterostructures\cite{Hu_2020_MachineLearningOptimizeda}.

The \glspl{ltc} of both perfect and disordered superlattices span a wide range, from approximately \qty{8}{\watt\per\meter\per\kelvin} to \qty{58}{\watt\per\meter\per\kelvin}. This demonstrates the promising potential for \gls{ltc} modulation in lateral heterostructures through ordered- or disordered-sequence engineering, particularly as the \gls{ltc} range of order- and disorder-sequence heterostructures is distinct (see \autoref{fig:TC}).

\begin{figure*}[htb]
\begin{center}
\includegraphics[width=2\columnwidth]{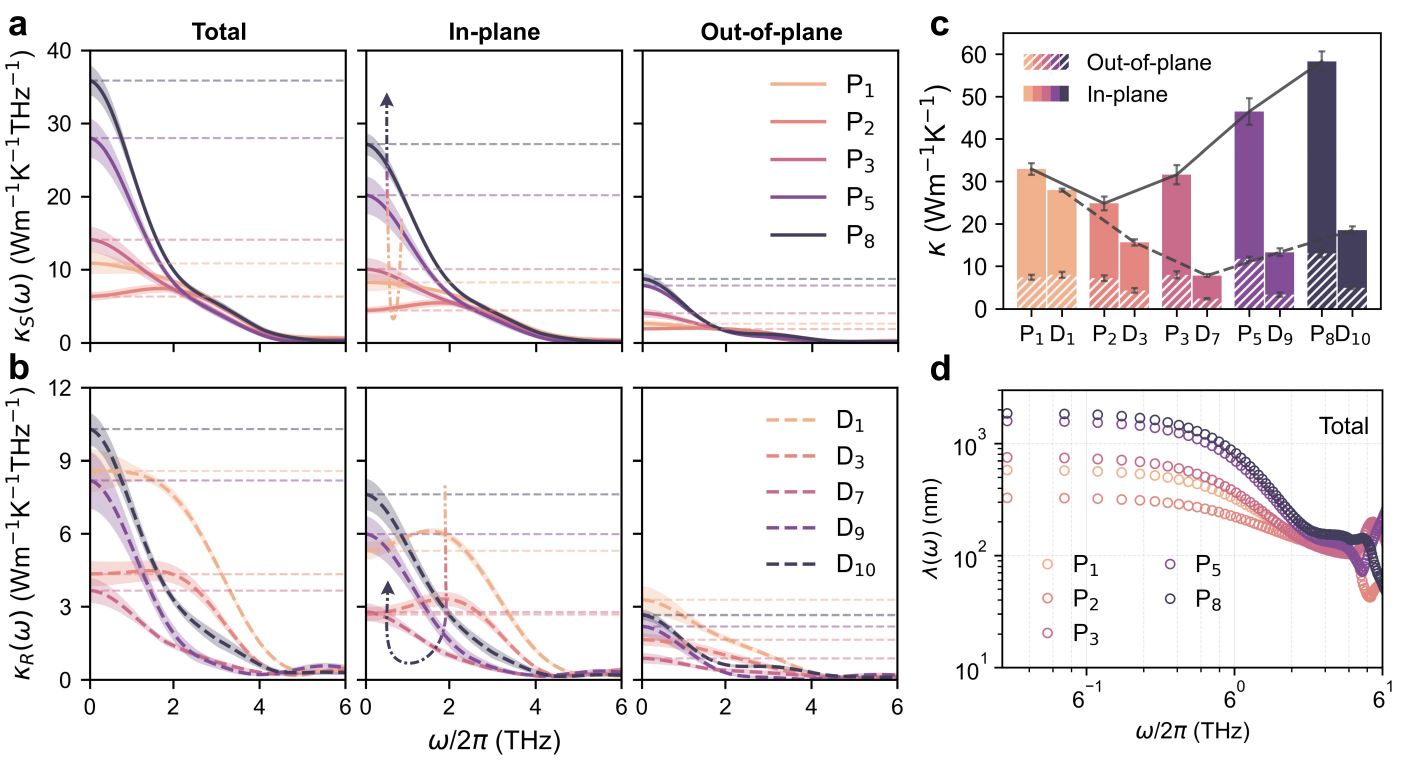}
\caption{Spectral \glspl{ltc} of selected MoSe$_2$/WSe$_2$ lateral heterostructures. (a-b) Total spectral \gls{ltc} (left column) and corresponding in-plane (middle column) and out-of-plane (right column) phonon contributions for (a) perfect and  (b) disordered superlattices. (c) Comparison of cumulative in-plane and out-of-plane phonon contributions to \gls{ltc} for each heterostructure. (d) Spectral phonon \gls{mfp} for five representative structures of the MoSe$_2$/WSe$_2$ lateral superlattice. In (a-b), the dashed lines with arrows indicate the trend of spectral \gls{ltc} peak that dominates the \gls{ltc} with the increasing interface density. The horizontal dashed lines are the reference lines for each $\kappa(\omega)$ value around \qty{0}{\tera\hertz}. Colored shaded areas represent error bars from multiple independent simulations.}
\label{fig:spectral}
\end{center}
\end{figure*}

To gain microscale insights into the interplay between particle-like and wave-like phonons in perfect and disordered superlattices, we spectrally decomposed the \glspl{ltc} within the \gls{hnemd} framework using the spectral heat current decomposition method (see \autoref{section:shc} for details)\cite{Fan_2019_Homogeneousa}. \autoref{fig:spectral}(a-b) present the spectral \gls{ltc} ($\kappa(\omega)$) results for perfect (P$_1$, P$_2$, P$_3$, P$_5$, and P$_8$) and disordered superlattices (D$_1$, D$_3$, D$_7$, D$_9$, and D$_{10}$), respectively, each with five representative pairs of perfect and disordered superlattices that have identical interface densities (see \autoref{fig:TC}), effectively capturing the changing trends. For perfect superlattices, the \glspl{ltc} are primarily contributed by low-frequency in-plane phonons below \qty{4}{\tera\hertz}. In the incoherent transport regime (P$_{1}$$\rightarrow$P$_{2}$), enhanced phonon-interface scattering significantly weakens \gls{ltc} from low-frequency phonons below \qty{4}{\tera\hertz}. Upon entering a coherent transport regime (P$_{2}$$\rightarrow$P$_{8}$), increasing phonon coherence enables more low-frequency, long-wavelength phonons to participate in transport, primarily concentrated in the in-plane part, also as shown in Figure \autoref{fig:spectral}(c). This is in contrast  to single-atom-thickness graphene/C$_3$N lateral heterostructure, where the \gls{ltc} is much higher and dominated by out-of-plane phonon modes \cite{Wu_2023_Dual}. This difference is attributed to the sandwich structure of MoSe$_{2}$ and WSe$_{2}$ makes their flexural phonon modes no longer the pure out-of-plane vibration. As a result, the symmetry selection rules are broken, allowing three-phonon scattering processes that involve an odd number of flexural phonons \cite{Gu_2018_Colloquium}.

As shown in \autoref{fig:spectral} (b), the $\kappa(\omega)$ of disordered superlattices, D$_1$ and D$_3$, with low interface density differs significantly from that of perfect superlattices, with peaks appearing around \qty{2}{\tera\hertz}, rather than near the \qty{0}{\tera\hertz} as in superlattices. Compared with D$_1$, the enhanced phonon-interface scattering in D$_3$ results in a weakened $\kappa(\omega)$ in the entire frequency range, with peaks attenuated but still present. However, when the interface density increases from around \qty{0.35}{\nano\meter^{-1}} (D$_3$) to \qty{1.76}{\nano\meter^{-1}} (D$_7$), the $\kappa(\omega)$ profile undergoes a complete transformation, with the peak shifting towards the \qty{0}{\tera\hertz}, resembling the peak shape observed in perfect superlattices. As indicated by the dashed arrows labeled in \autoref{fig:spectral}(a-b), disordered superlattices show spectral peak shifts that are absent in superlattices. 
In the high interface density region (D$_7$$\rightarrow$D$_{10}$) where \gls{ltc} increases with interface density, in-plane phonons below \qty{2}{\tera\hertz} dominants the increase and exhibit coherent characteristics similar to those observed in perfect superlattices (also see P$_2$$\rightarrow$P$_{8}$ in \autoref{fig:spectral}(c)). These results suggest that phonon coherence contributes to the increase in \gls{ltc} in disordered superlattices at high interface densities, as the smaller differences between adjacent units in these configurations facilitate the transmission of low-frequency, long-wavelength wave-like phonons.

As defined in Eq.~\eqref{equation:kw} (see \autoref{section:hnemd}), one can predict frequency-dependent phonon \gls{mfp}, $\lambda(\omega)$, using the following equation:\cite{Fan_2019_Homogeneousa}
\begin{equation}
   \lambda(\omega)=\frac{\kappa(\omega)}{G(\omega)},
\end{equation}
where $G (\omega)$ is the spectrally thermal conductance in the ballistic regime. \autoref{fig:spectral}(d) depicts the $\lambda(\omega)$ profiles for representative perfect superlattices below \qty{6}{\tera\hertz}, the frequency range that contributes most to \gls{ltc}. The phonon \glspl{mfp} align with their \glspl{ltc}, especially for phonons below \qty{2}{\tera\hertz}. In the coherent transport regime, enhanced phonon coherence reduces scattering and extends the phonon \glspl{mfp}. Coherent phonon wave packets maintain their phase relationships, minimizing incoherent scattering, particularly for low-frequency phonons, whose longer wavelengths favor coherent states.

\section{Summary and conclusions}
\label{section:summary}

In summary, we develop an accurate and efficient machine-learned \gls{nep} model for lateral MoSe$_2$/WSe$_2$ heterostructure, and then perform extensive \gls{hnemd} simulations to investigate thermal transport in both perfect and disordered MoSe$_2$/WSe$_2$ superlattices. Surprisingly, both perfect and disordered superlattices show a minimum \gls{ltc} with varying interface density, driven by the transition from incoherent to coherent transport regimes. The counter-intuitive thermal transport behavior of disorder superlattices is well captured by the disordering factor, defined by the auto-correlation function of the parameterized sequence. For the first time, spectral \gls{ltc} decomposition reveals wave-like low-frequency (long-wavelength) phonons contribute to the increase in \gls{ltc} of disordered superlattices with high interface densities. We also explained the intrinsic mechanism of coherent phonons enhancing \gls{ltc} in the perfect superlattices through phonon \gls{mfp} and stop bands analysis. These findings provide novel physical insights into tuning thermal transport of \gls{2d} lateral heterostructures through sequence ordering and disordering, offering a framework applicable to other analogous systems.

\section{methods}
\subsection{The training dataset}
\label{section:dataset}

As shown in \autoref{fig:nep}(a), our training dataset consists of nine different configurations, i.e., two pure MX$_2$ structures (MoSe$_2$ and WSe$_2$), two MoSe$_2$/WSe$_2$ heterostructures with the ideal flat interface and mutually embedded interface, and five random ternary transitional Mo$_x$W$_{1-x}$Se structures with $x$ increasing from $\frac{1}{6}$ to $\frac{5}{6}$ in increments of $\frac{1}{6}$. For each configuration, we performed constant volume \gls{md} simulation driven by the \gls{sw} potential developed by Jiang \textit{et al.}~\cite{jiang2019misfit} for 150 ps with the target temperature linearly increasing from 100 to 800 K  and sampled 15 structures. In addition to this, we  generated 5 structures by applying random cell deformations (-3 to 3\%) and atomic displacements (within 0.1 $\mathrm{\AA}$) starting from each initial configuration. In total, we obtained $9\times(15+5)=180$ structures, which were further randomly divided to  a training dataset (150 structures) and a test dataset (30 structures). 

\subsection{DFT calculations}
\label{section:dft}
To obtain the total energy, atomic forces, and virial for selected structures in the reference dataset, we use the \textsc{VASP} code \cite{kresse1996efficiency, Kresse1996PRB, Kresse1999PRB} to perform single-point \gls{dft} calculations at PBE \cite{blochl1994projector} level. A plane-wave basis set was employed with the energy cutoff of 650 eV and a dense $\Gamma$-centered grid with a $k$-point density of 0.25/$\mathrm{\AA}$ was sampled in the Brillouin zone. We set a threshold of 10$^{-7}$ eV for the electronic self-consistent loop. 

\subsection{NEP model training}
\label{section:nep}
Using the provided training and test datasets, we trained the \gls{nep} model for the MoSe$_2$/WSe$_2$ heterosystem with the NEP3 architecture, implemented in \textsc{GPUMD}.\cite{Fan_2022_GPUMD} The \gls{nep} approach \cite{Fan_2021_Neuroevolution,Fan_2022_Improvinga,Fan_2022_GPUMD} uses a simple feedforward \gls{ann} to represent the site energy $U_{i}$ \cite{behler2007prl} of atom $i$ as a function of a descriptor vector with $N_{\mathrm{des}}$ components:
\begin{equation}
    U_i = \sum_{\mu=1}^{N_{\mathrm{neu}}} w _ \mu ^{(1)} \tanh\left(\sum_{\nu=1}^{N_{\mathrm{des}}}w_{\mu\nu}^{(0)}q_\nu^i-b_\mu^{(0)}\right)-b^{(1)},
\end{equation}
where $\tanh(x)$ is the activation function, $w^{(0)}$, $w^{(1)}$, $b^{(0)}$, and $b^{(1)}$ are the trainable weight and bias parameters in the \gls{ann}, and $q^{i}_{\nu}$ is the descriptor vector constructed similarly to the \gls{ace} approach \cite{drautz2019atomic}, but treating radial and angular components separately. Based on the \gls{snes} method~\cite{Schaul2011}, the optimization of \gls{nep} model parameters involves minimizing a loss function that incorporates a weighted sum of the \glspl{rmse} for energy, force, and virial, along with regularization terms. 

After extensive testing, we determined the following hyper-parameters for \gls{nep} training: the cutoffs of radial and angular descriptors were both set to \qty{5}{\angstrom} and with 8 radial functions. For angular terms, we focused on three-body and four-body interactions, with maximum expansion orders of 4 and 2, respectively. A single hidden layer with 50 neurons was used for the \gls{ann}. The optimization was performed with a population size of $50$ for \num{5e5} steps. In the loss function, we assigned weights of 1.0 to the \glspl{rmse} of energy and force, and 0.1 to the \gls{rmse} of virial.

\subsection{The HNEMD method}
\label{section:hnemd}
Based on the linear response theory, the \gls{hnemd} method simulates the thermal gradient effect in a solid by applying a directional driving force $\boldsymbol{F}_{i}^{\mathrm{e}}$ on each atom $i$ \cite{Evans_1982_Homogeneousb, Fan_2019_Homogeneousa}:
\begin{equation}
\label{equation:fe}
    \boldsymbol{F}_{i}^{\mathrm{e}} = \boldsymbol{F}_{\mathrm{e}} \cdot \mathbf{W}_{i},
\end{equation}
where $\boldsymbol{F}_\mathrm{e}$ is the external driving force parameter with the dimension of inverse length, and $\mathbf{W}_{i}$ is virial tensor of atom $i$. 
The running \gls{ltc}, $\kappa(t)$, along the lateral transport direction can be calculated as:
\begin{equation}
\label{equation:kappa}
\kappa(t)=\frac{\langle J(t)\rangle_{\mathrm{ne}}}{T V F_{\mathrm{e}}},
\end{equation}
where $k_{\mathrm{B}}$ is the Boltzmann constant, $T$ is the temperature, $V$ is the volume of the system, and $\left\langle \boldsymbol{J} \right\rangle_{\mathrm{ne}}$ is non-equilibrium ensemble average of the heat current $\boldsymbol{J} = \sum_{i}\mathbf{W}_{i}\cdot\boldsymbol{v}_{i}$, with $\boldsymbol{v}_{i}$ being the velocity of atom $i$.
A thickness of \qty{0.65}{\nano\meter} is used to calculate the volume $V$ in Eq.~\eqref{equation:kappa} for all structures.
To observe the convergence of the calculated \gls{ltc} results, Eq.~\eqref{equation:kappa} can be further redefined as the following cumulative average: $\kappa(t) = \frac{1}{t}\int_{0}^{t}\kappa(\tau) \mathrm{d}\tau$. 

We set the magnitude of the driving force parameter to be $F_{\mathrm{e}}=$\qty{1e-5}{\per\angstrom} along lateral transport directions for all heterostructures, which is small enough to maintain the linear response regime and is large enough to achieve a sufficiently large signal-to-noise ratio. All heterostructures were set to a length of \qty{56.95}{\nano\meter} along the thermal transport direction, with a width of \qty{9.86}{\nano\meter} and a total of 18,000 atoms. Periodic boundary conditions were applied in both in-plane directions. For the superlattice group, five independent runs were conducted to enhance the statistical accuracy and obtain an error estimate. For the disordered group, three different configurations were created for each interface density, with three independent simulations performed for each configuration. All \gls{md} simulations were performed using \textsc{gpumd} package.\cite{Fan_2022_GPUMD}. In all \gls{md} simulations, the system was first equilibrated at \qty{300}{\kelvin} and zero pressure for \qty{1}{\nano\second} in the NPT ensemble. In the production stage, the Nose-Hoover chain thermostat \cite{tuckerman2023statistical} was used to maintain the overall temperature and heat current data were collected for \qty{10}{\nano\second}. A time step of \qty{1}{\femto\second} was used in all the \gls{md} simulations. 

\subsection{Spectral heat current decomposition}
\label{section:shc}
In the framework of the \gls{hnemd} method, one can calculate spectrally decomposed thermal conductivity with the following formula \cite{Fan_2019_Homogeneousa}:
\begin{equation}
\kappa = \int_0^{\infty} \frac{d\omega}{2\pi} \kappa(\omega),
\end{equation}
where
\begin{equation}
\label{equation:kw}
\kappa(\omega) = \frac{2}{VTF_{\rm e}} \int_{-\infty}^{\infty} dt e^{i\omega t} K(t).
\end{equation}
Here, $K(t)$ is the $x$-component (along lateral direction) of the virial-velocity correlation function \cite{gabourie2021spectral}:
\begin{equation}
\bm{K}(t) = \sum_i \langle \mathbf{W}_i(0) \cdot \bm{v}_i(t) \rangle.
\end{equation}
For \gls{2d} materials considered here, the spectral \gls{ltc} $\kappa(\omega)$ in Eq.~\eqref{equation:kw} can be further decomposed into in-plane and out-of-plane (flexural) phonon contributions \cite{fan2017thermal}.

\begin{acknowledgments}
X. Wu is the JSPS Postdoctoral Fellow for Research in Japan (No. P24058). 
T. Liang and J. Xu acknowledge support from the National Key R\&D Project from the Ministry of Science and Technology of China (Grant No. 2022YFA1203100), the Research Grants Council of Hong Kong (Grant No. AoE/P-701/20), and RGC GRF (Grant No. 14220022).
X. Wu and M. Nomura acknowledge support from the JSPS Grants-in-Aid for Scientific Research (Grant Nos. 21H04635) and JST SICORP EIG CONCERT-Japan (Grant No. JPMJSC22C6).
P. Ying is supported by the Israel Academy of Sciences and Humanities \& Council for Higher Education Excellence Fellowship Program for International Postdoctoral Researchers. 
\end{acknowledgments}

\noindent{\textbf{Conflict of Interest}}

The authors have no conflicts to disclose.

\ 
\

\noindent{\textbf{Data availability}}

The source code and documentation for  \textsc{gpumd} are available
at \url{https://github.com/brucefan1983/GPUMD} and \url{https://gpumd.org}, respectively. The documentation for \textsc{calorine} is available
at \url{https://calorine.materialsmodeling.org}.
The documentation for \textsc{phonopy} is available
at \url{https://phonopy.github.io/phonopy/}. 
The inputs and outputs related to the NEP model training are freely available at the Gitlab repository \url{https://gitlab.com/brucefan1983/nep-data}.

\bibliography{refs}

\end{document}